\documentclass[useAMS,usenatbib,aas_macros]{mn2e}

\usepackage{amssymb,amsmath}
\usepackage{graphicx}
\usepackage{color}
\usepackage{flafter}
\usepackage{placeins}
\usepackage{subfigure}
\usepackage{booktabs}
\usepackage{array}
\usepackage{natbib}
\usepackage{aas_macros}
\usepackage{multirow}
\usepackage[T1]{fontenc} 
\usepackage{aecompl}
\usepackage{float}
\usepackage[draft]{hyperref} 

\title[Boxy/peanut bulges and gas inflow]{A close look at secular evolution:\newline Boxy/Peanut bulges reduce gas inflow to the central kiloparsec}
\author[Fragkoudi et al.]{F. Fragkoudi$^{1,2}$\thanks{E-mail:
francesca.fragkoudi@obspm.fr}, E. Athanassoula$^{1}$, A. Bosma$^{1}$  \\
$^{1}$Aix Marseille Universit\'{e}, CNRS, LAM (Laboratoire d'Astrophysique de Marseille) UMR 7326, 13388, Marseille, France \\
$^{2}$GEPI, Observatoire de Paris, CNRS, Universit\'{e} Paris Diderot, 5 Place Jules Janssen, 92195
Meudon, France \\}
\begin{document}

\date{}

\pagerange{\pageref{firstpage}--\pageref{lastpage}} \pubyear{2014}

\maketitle

\label{firstpage}

\begin{abstract}
In this letter we investigate the effect of boxy/peanut (b/p) bulges on bar-induced gas inflow to the central kiloparsec, which plays a crucial role on the evolution of disc galaxies.
We carry out hydrodynamic gas response simulations in realistic barred galaxy potentials, including or not the geometry of a b/p bulge, to investigate the amount of gas inflow induced in the different models. We find that b/p bulges can reduce the gas inflow rate to the central kiloparsec by more than an order of magnitude, which leads to a reduction in the amount of gas available in the central regions. We also investigate the effect of the dark matter halo concentration on these results, and find that for maximal discs, the effect of b/p bulges on gas inflow remains significant. The reduced amount of gas reaching the central regions due to the presence of b/p bulges could have significant repercussions on the formation of discy- (pseudo-) bulges, on the amount of nuclear star formation and feedback, on the fuel reservoir for AGN activity, and on the overall secular evolution of the galaxy.

\end{abstract}




\section{Introduction}

Non-axisymmetric structures such as bars -- which are found in about two thirds of disc galaxies in the local universe \citep{Eskridgeetal2000,Menendezetal2007,Aguerrietal2009,Gadotti2009,Butaetal2010,Butaetal2015} -- are thought to be the main driver of the secular phase of galaxy evolution. They induce torques which force the gas to shock, subsequently losing angular momentum and funneling to the centre. This contributes to shaping the central regions of disc galaxies by creating discy-pseudobulges \citep{KormendyKennicutt2004,Athanassoula2005}, triggering nuclear star formation \citep{Ellisonetal2011} and feedback, and possibly fuelling AGN activity (\citealt{Shlosmanetal1990,CoelhoGadotti2011, Emsellemetal2014}). 

Boxy/peanut/X-shaped (b/p) bulges (also referred to as boxy/peanuts, or b/ps) are structures which extend out of the plane of nearly half of all edge-on disc galaxies in the local universe \citep{Luttickeetal2000}. Numerical simulations and orbital structure analysis have shown the intimate link between bars and b/ps, by demonstrating that b/ps are caused by vertical orbital instabilities in bars, which cause the bar to puff out vertically from the disc (e.g. \citealt{Binney1981,PfennigerFriedli1991,Skokosetal2002}) and that once a bar forms a b/p bulge will form soon after (e.g. \citealt{Combesetal1990,MartinezValpuestaetal2006}.) For reviews on these topics the reader is referred to \cite{Athanassoula2008,Athanassoula2016} and references therein. Additionally, kinematic studies of edge-on barred galaxies and b/ps also confirm the connection between the two structures \citep[and references therein]{Athanassoula&Bureau1999, Bureau&Athanassoula1999, Chung&Bureau2004, Bureau&Athanassoula2005}. Furthermore, photometric and kinematic studies have shown that the bulge of our own Milky Way also contains such a b/p \citep{Weilandetal1994, Howardetal2009, NessLang2016}, and indeed recent results suggest that the \emph{main} component of the Milky Way bulge is the b/p, with little room left for a massive classical bulge (\citealt{Shenetal2010}, Ness et al. 2012, 2013; \nocite{Nessetal2012,Nessetal2013a} \citealt{WeggGerhard2013, DiMatteoetal2014}).

Although the effect of bars on gas inflow has been extensively examined in the literature (e.g. \citealt{Athanassoula1992b,Pineretal1995,ReganTeuben2004,Kimetal2012,Lietal2015}), a study of the effects of the b/p geometry on the gas inflow has not -- until present -- been carried out. Recent work on the effects of b/p bulge geometry on galaxy models showed that b/ps have a significant effect on the bar strength and orbital structure of barred galaxies, and as such hints at the possible effect these structures could have on the gas inflow \citep{Fragkoudietal2015}. A detailed study of the effect of b/ps on the gas dynamics of spiral galaxies is therefore needed, if we are to fully understand the effects of secular evolution on gas transportation to the centre.

As bar-induced disturbances are due to a non-linear response of the gas to the gravitational potential, they are best studied using numerical techniques, such as hydrodynamic simulations. We therefore carry out hydrodynamic gas response simulations in models with and without b/ps, but which are otherwise identical, in order to examine the effects of b/ps on the gas inflow to the central kiloparsec.\footnote{It is worth noting that we do not add any additional mass when modelling the b/p bulge; we simply redistribute the mass of the disc in a geometry representative of the b/p, thus changing the volume density of the model. Therefore, in what follows we interchangeably use the terms ``geometry of the b/p bulge'' and simply ``b/p bulge''.}
In Section \ref{sec:5models} we present the models used in this study, in Section \ref{sec:results} we present the results, and in Section \ref{sec:5summary} we summarise and give some concluding remarks.

\section{The simulations}
\label{sec:5models}

\begin{figure}
\centering
\includegraphics[width=0.9\linewidth]{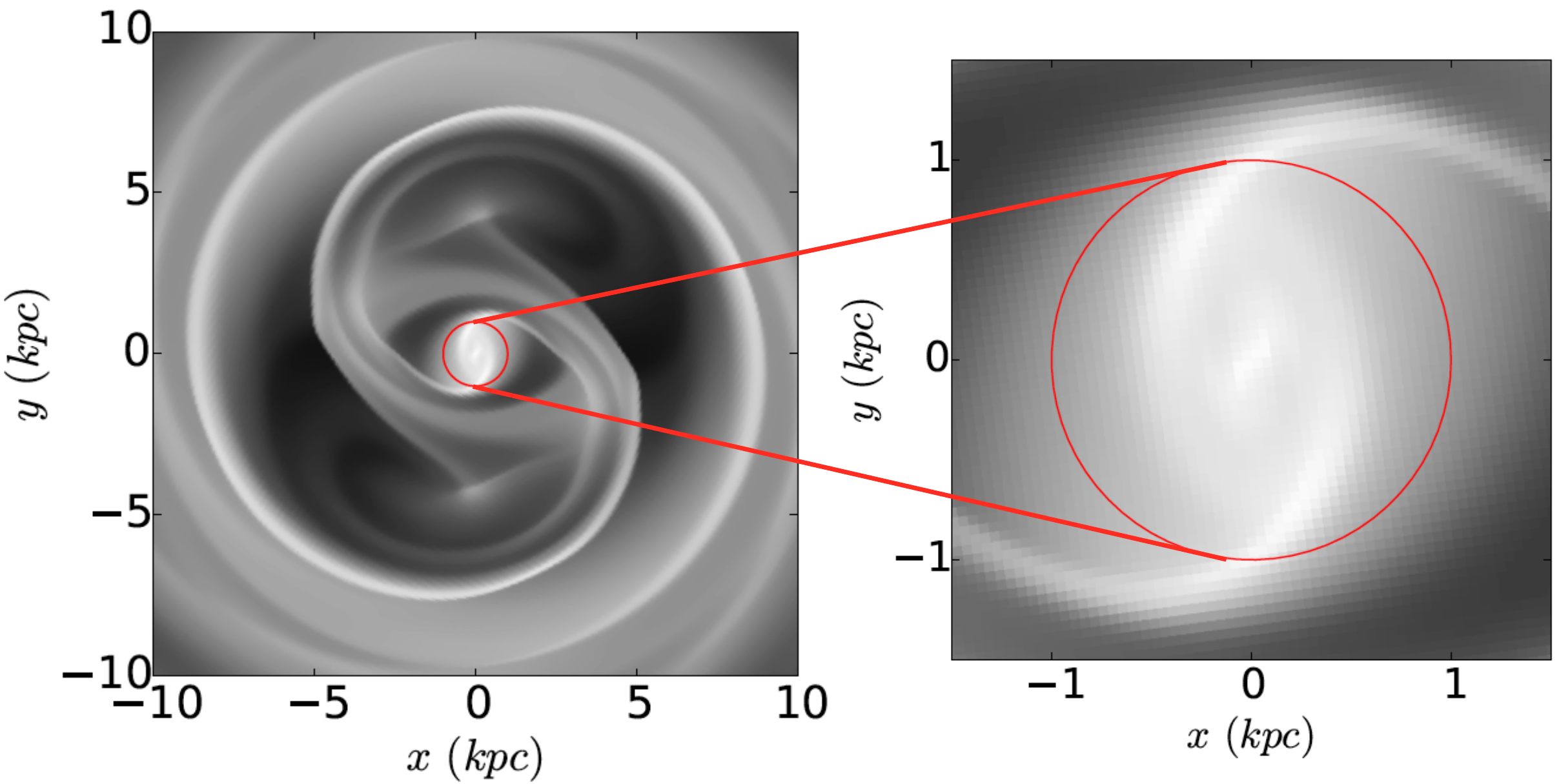}
\caption{\emph{Left:} Gas surface density for model \emph{noBP}, with a red line denoting a 1\,kpc circle, within which we consider the gas inflow. \emph{Right:} A zoom in of the region around 1.5\,kpc.} 
\label{fig:3surfdens}
\end{figure}

\begin{figure}
\centering
\includegraphics[width=1.\linewidth]{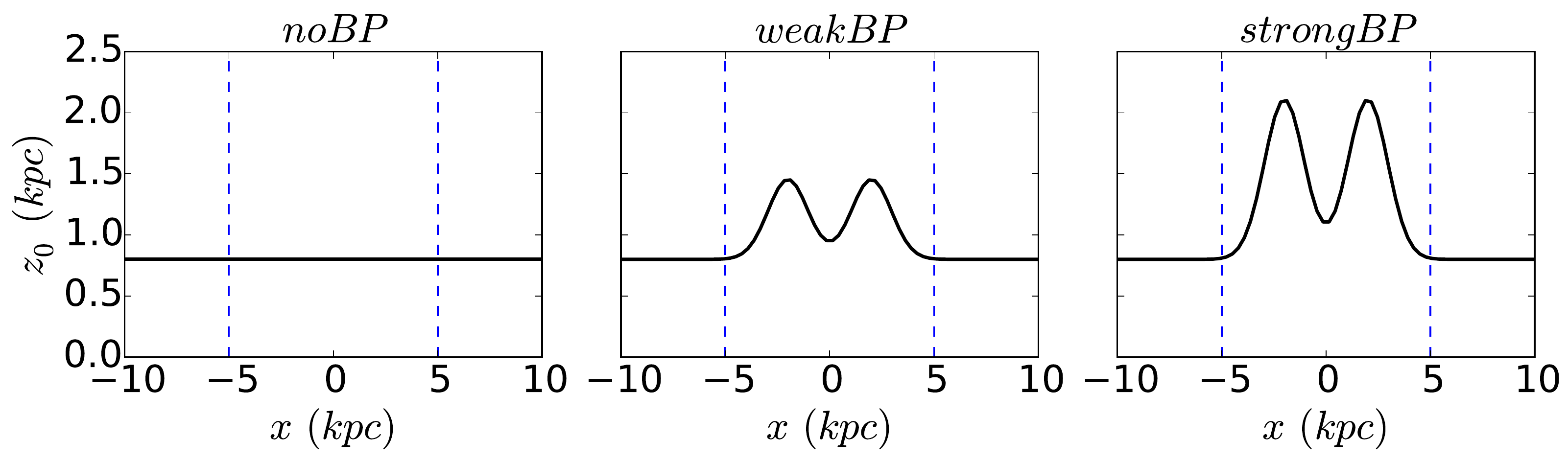}
\caption{Scaleheight ($z_0$) along the $x$-axis for the three models used in this study. From left to right: model \emph{noBP} - the scaleheight is constant throughout the disc, model \emph{weakBP} - the model contains a relatively weak b/p, and model \emph{strongBP} - a model with a strong b/p. The dashed blue lines indicate where the bar ends.} 
\label{fig:z0_3}
\end{figure}

\begin{figure}
\centering
\includegraphics[width=0.7\linewidth]{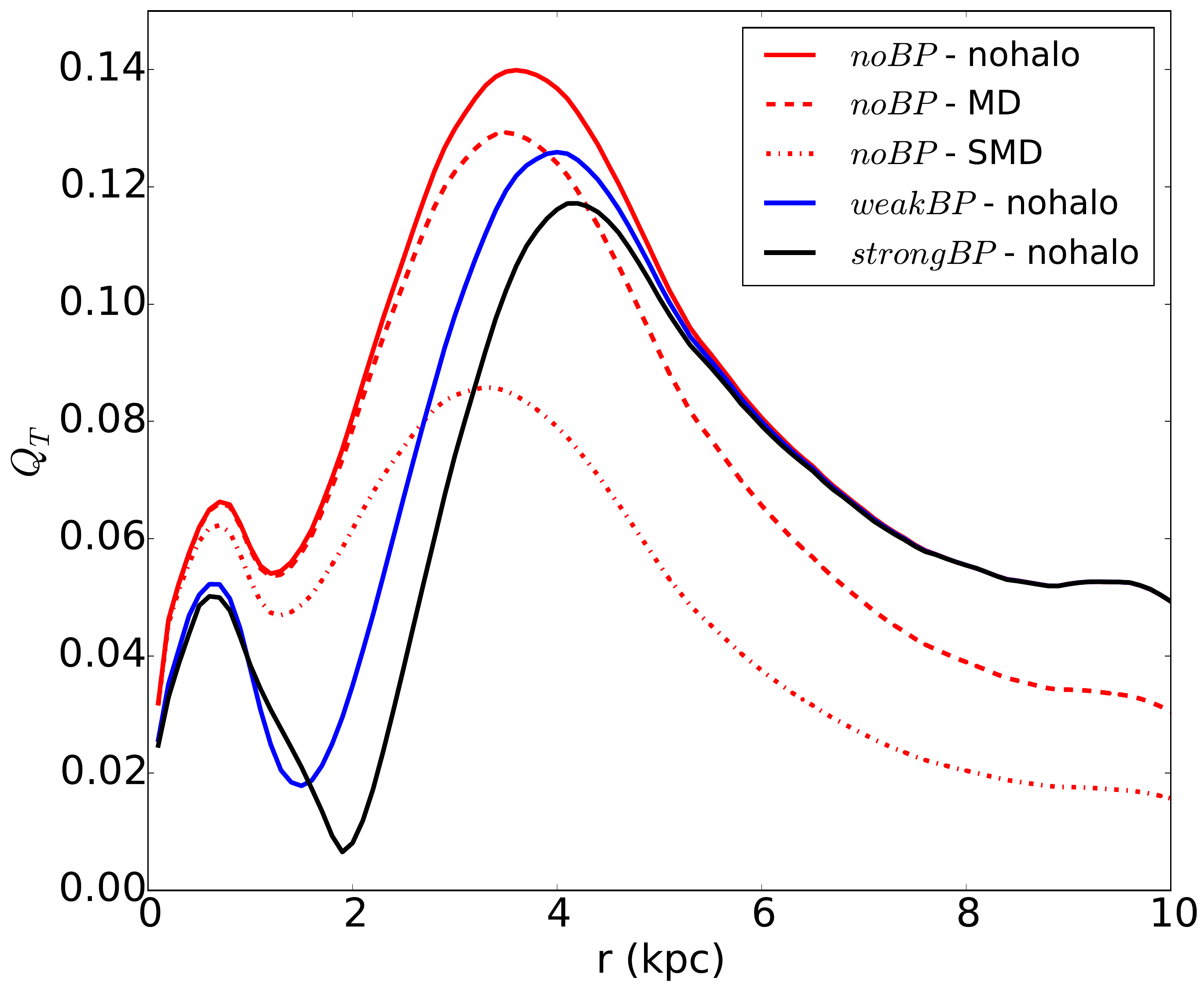}
\caption{The bar strength, as given by $Q_T$ -- the maximum of the tangential forces divided by the azimuthal average of the radial forces -- for some of the models considered here. We see that both adding a b/p bulge and adding a dark matter halo reduce the bar strength.} 
\label{fig:qt}
\end{figure}

We construct realistic potentials of a barred galaxy from near-infrared (NIR) images of NGC 1291 -- an early type S0/a barred galaxy -- taken from the S$^4$G survey \citep{Sheth2010_s4g} in which we run hydrodynamic gas response simulations (see Figure \ref{fig:3surfdens}). NIR images are best suited for obtaining the potential of galaxies, since they trace the old stellar population, where most of the stellar mass of the galaxy is. We use NGC 1291 for this study, since it is a face-on barred galaxy and therefore we do not need to deal with deprojection effects. The bar has a semi-major axis of $\approx$5\,kpc while the galaxy also has what appears to be a barlens, indicating the presence of a b/p bulge \citep{Laurikainenetal2014,Athanassoulaetal2015}.
The models of NGC 1291 are obtained by assigning a Mass-to-Light ratio (M/L) = 0.6 $M_{\odot}$/$L_{\odot}$ \citep{Meidtetal2014,Roecketal2015, Querejetaetal2015} and then ascribing a height function to the model.

We focus on three different models, one without a b/p, one with a weak and one with a strong b/p bulge. These models have a height function given by,
\begin{equation}
F(x,y,z)=\frac{1}{2 z_0(x,y)}\mathrm{sech}^2\left(\frac{z}{z_0(x,y)}\right),
\label{eq:peanut1}
\end{equation}
where $z_0$ is the scaleheight. For the model without a b/p, (model \emph{noBP}) $z_0$ is a constant, while for the models with a weak and strong b/p (dubbed \emph{weakBP} and \emph{strongBP} respectively) the scaleheight varies along the disc as the sum of two two-dimensional gaussians (see Figure \ref{fig:z0_3}).

These models all have an outer disc scaleheight of $z_0$=0.8kpc; the \emph{weakBP} and \emph{strongBP} models have a peanut height function with a maximum scaleheight of the peanut of 1.45 and 2.1\,kpc respectively. These values are chosen by fitting the height function to a N-body simulation of a barred galaxy containing a b/p (for more details on the height function and the N-body simulations the reader is referred to \citealt{Fragkoudietal2015} and \citealt{Athanassoulaetal2013}).
It is important to emphasise that \emph{all} the other parameters in the models are the same -- the only difference between the models is the strength of the b/p. As can be seen in Figure \ref{fig:qt} (solid curves), adding a b/p significantly reduces the bar strength given by $Q_T$ \citep{CombesSanders1981,ButaBlock2001}, i.e. the maximum of the ratio of tangential forces to azimuthally averaged radial forces (see also \citealt{Fragkoudietal2015}).

In Section \ref{sec:dmflow} we add a dark matter halo to the models, in order to study the effect of the halo concentration on the results presented in this study. The dark matter halo is modelled as a pseudo-isothermal sphere and its parameters are adjusted such that the outer flat part of the rotation curve matches that predicted by the Baryonic Tully-Fisher relation for NGC 1291 \citep{McGaughetal2000,Zaritskyetal2014}. 
We consider two different cases: a model with a maximal disc and a non-concentrated dark matter halo (we use the maximal disc definition by \citealt{Sackett1997}, i.e. the stellar rotation curve is 75$\pm$10\% the total rotation curve at 2.2 disc scalelengths), denoted by MD, and a model with a sub-maximal disc and a concentrated dark matter, denoted by SMD. We see in Figure \ref{fig:qt} (red curves) that adding a dark matter halo also decreases the bar strength of the models. 

We run two-dimensional hydrodynamic simulations in all these potentials and examine the gas inflows to the central kiloparsec (see Figure \ref{fig:3surfdens}). The simulations are run with the adaptive mesh refinement code \textsc{ramses} \citep{Teyssier2002}.
The initial conditions of all the simulations are an axisymmetric disc in hydrostatic equilibrium with the same mass as the non-axisymmetric model (i.e. the mass of the model is constant throughout the duration of the simulation). The non-axisymmetric component is introduced gradually, from 0.2 to 0.9\,Gyr in order to avoid transients. The time over which the non-axisymmetric potential is grown, was found empirically as $\sim$3 bar rotations, which is enough in order to converge to a stable model, while not being excessively penalising in computing time; this is in accordance with previous studies \citep{Athanassoula1992b,PatsisAthanassoula2000}. The pattern speed is set such that the corotation radius falls just outside the bar semi-major axis. The gas is modelled as an infinitely thin disc of an isothermal perfect fluid with a sound speed of 10\,km/s, which is a reasonable value for the interstellar medium of galaxies. The resolution of the simulation is 50\,pc per grid size everywhere and the size of the box is 50\,kpc per side. We carried out tests to examine the impact of increasing the resolution of the grid and found that the conclusions of the study remain the same.

\begin{figure*}
\centering
\subfigure[Inflow rate]{%
	\includegraphics[height=6.25cm]{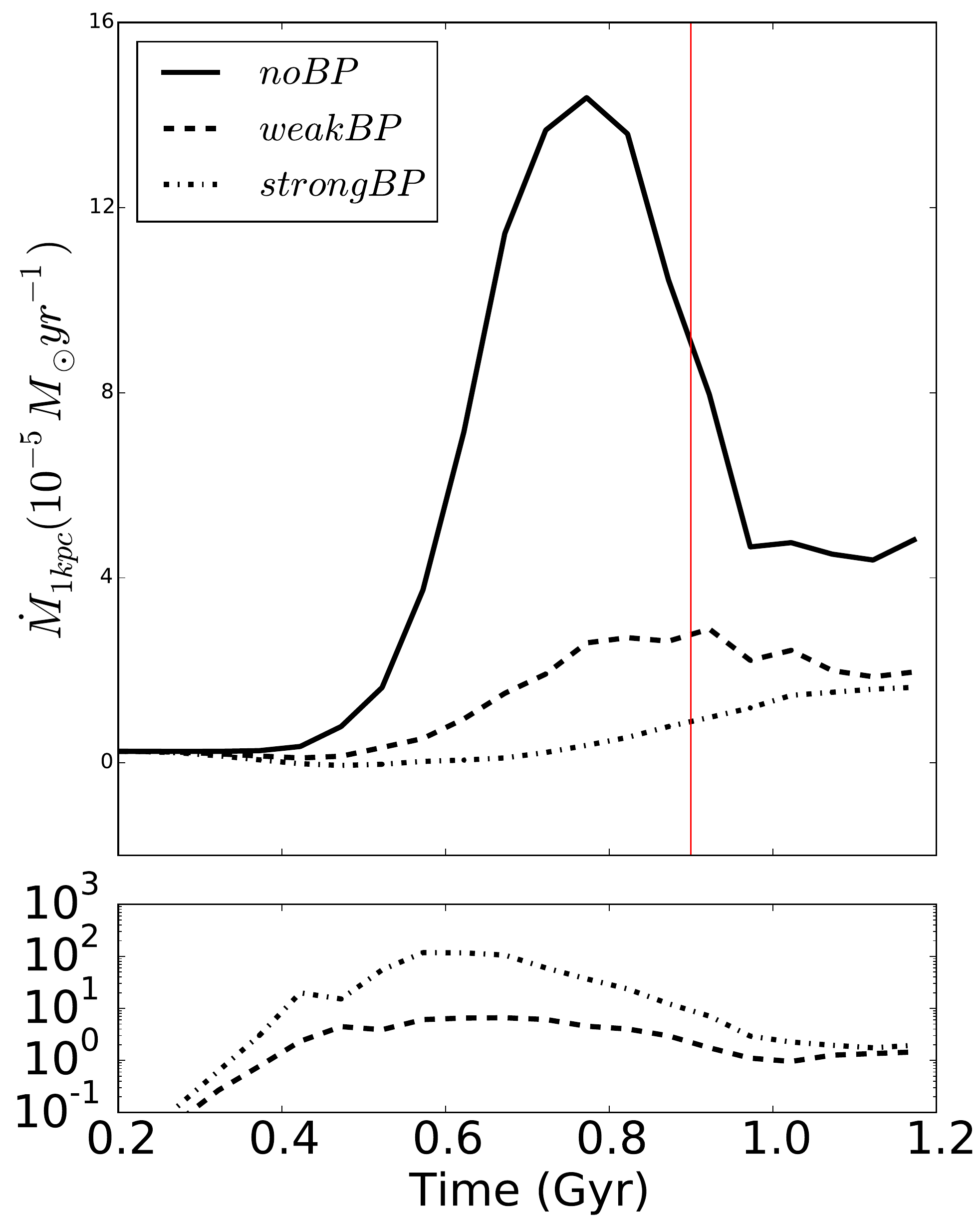}
	\label{fig:rate}}
\quad
\subfigure[Total gas mass]{%
	\includegraphics[height=6.25cm]{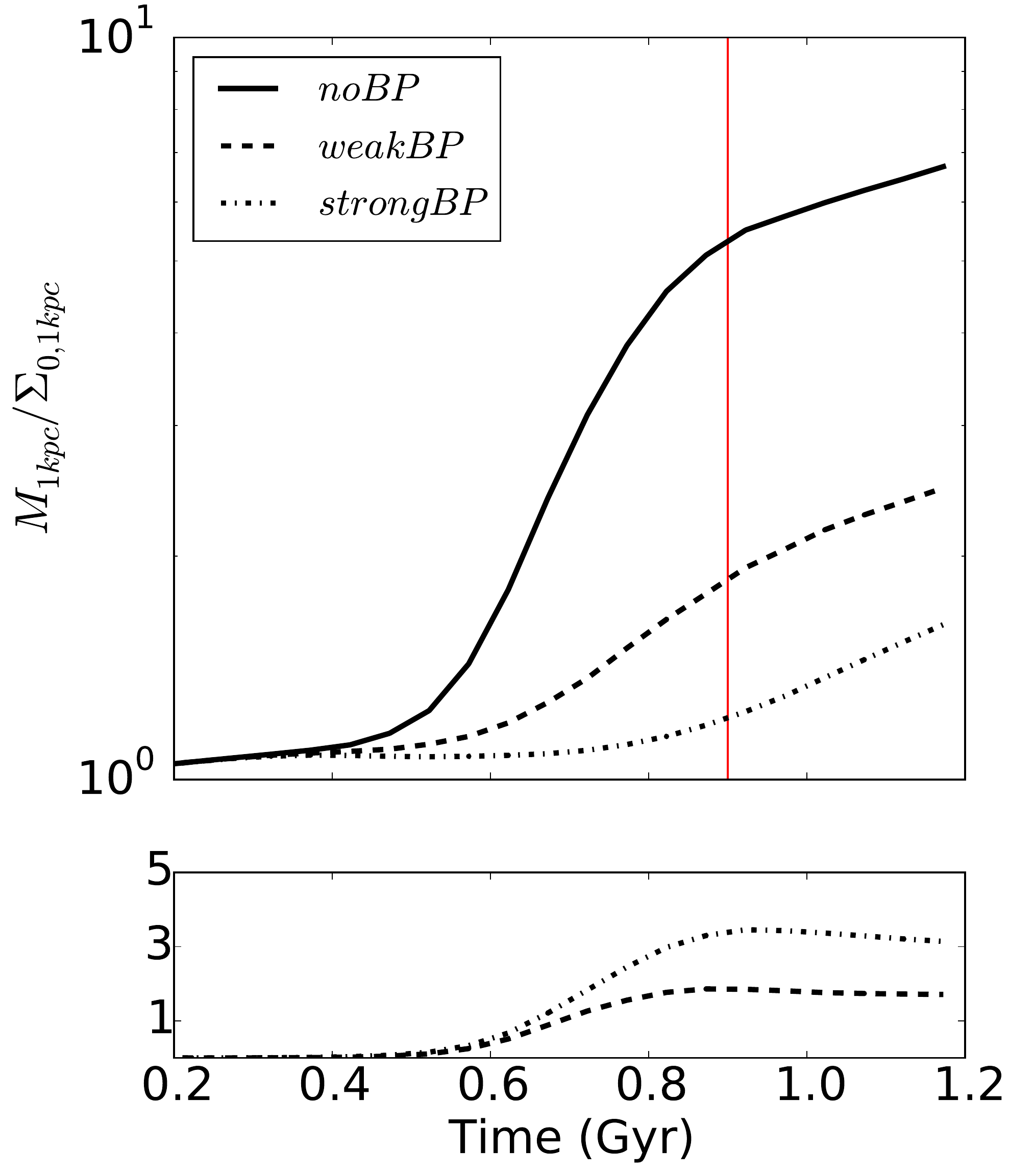}
	\label{fig:tot}}
\quad
\subfigure[Toy model of gas inflow]{%
	\includegraphics[height=6.25cm]{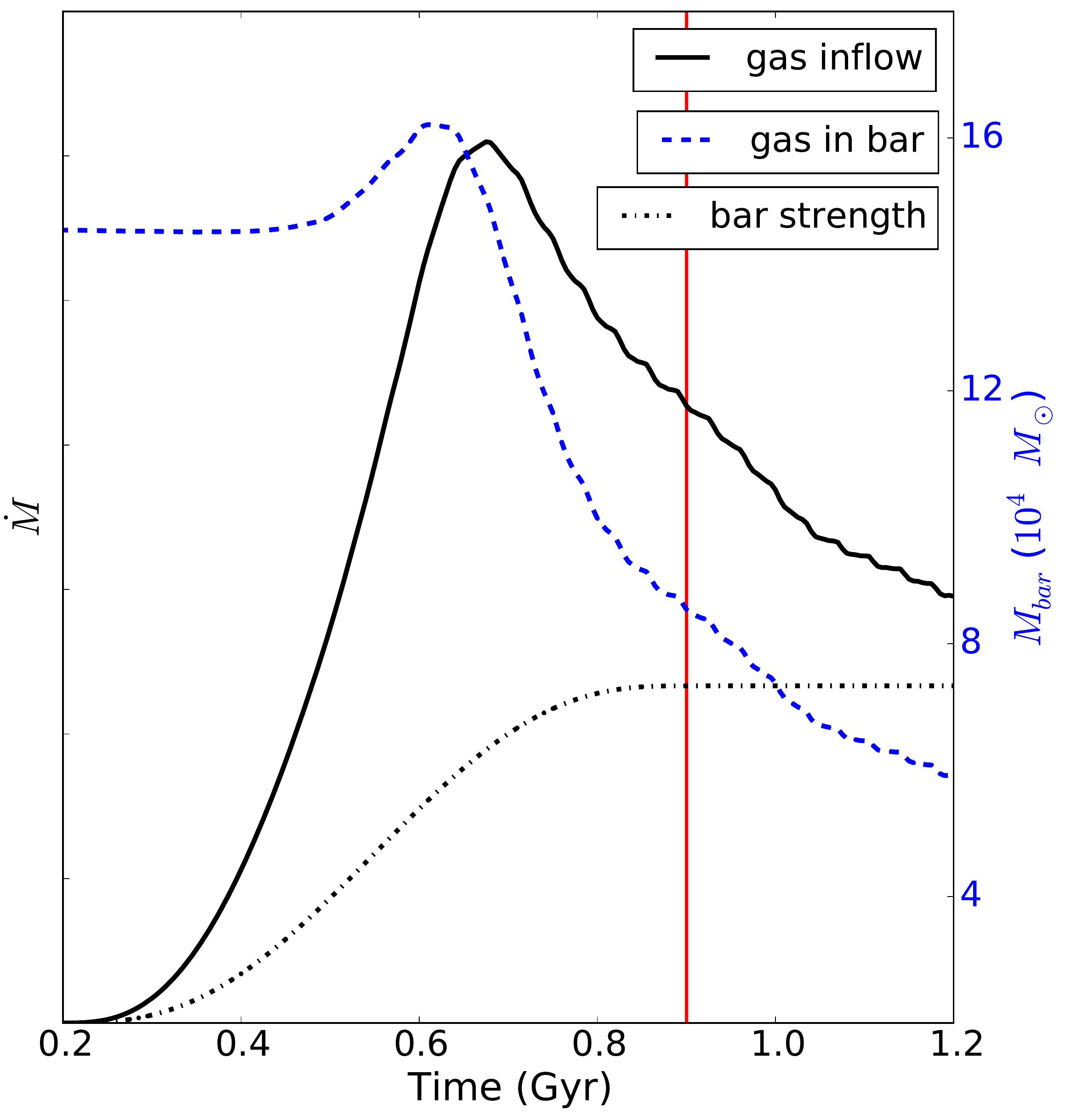}
	\label{fig:rate_norm}}
\quad
\caption{\emph{Left and Middle:} The gas inflow rate and total gas mass (normalised by the initial surface density) within a radius of 1\,kpc as a function of time, for the three models as indicated in the inset. The bottom panels give the relative difference, i.e. $abs((X_{noBP}-X_{weakBP, strongBP})/X_{weakBP, strongBP}$), between $weakBP$ and $noBP$ (dashed line) and $strongBP$ and $noBP$ (dot-dashed line), where $X$ stands for $\dot{M}$ and $M_{1kpc}$. We see that the gas inflow rate and total gas mass in the models with the weak and strong b/p are considerably reduced compared to the model without the b/p bulge. \emph{Right:} Toy model of gas inflow: the gas inflow rate in arbitrary units (solid) is obtained by multiplying the gas available in the bar region (dashed) with the bar strength (dot-dashed) times an arbitrary multiplicative constant. We see that the toy model approximately reproduces the curve for the gas inflow within 1\,kpc. In all plots the vertical red line indicates when the non-axisymmetric potential has been fully introduced, at 0.9\,Gyr.} 
\label{fig:rateandtot}
\end{figure*}
\section{Results}
\label{sec:results}

\subsection{The effect of b/p geometry on gas inflow}
\label{sec:gasinflow}

\begin{figure}
\centering
\subfigure[Maximal Disc]{%
	\includegraphics[width=0.45\linewidth]{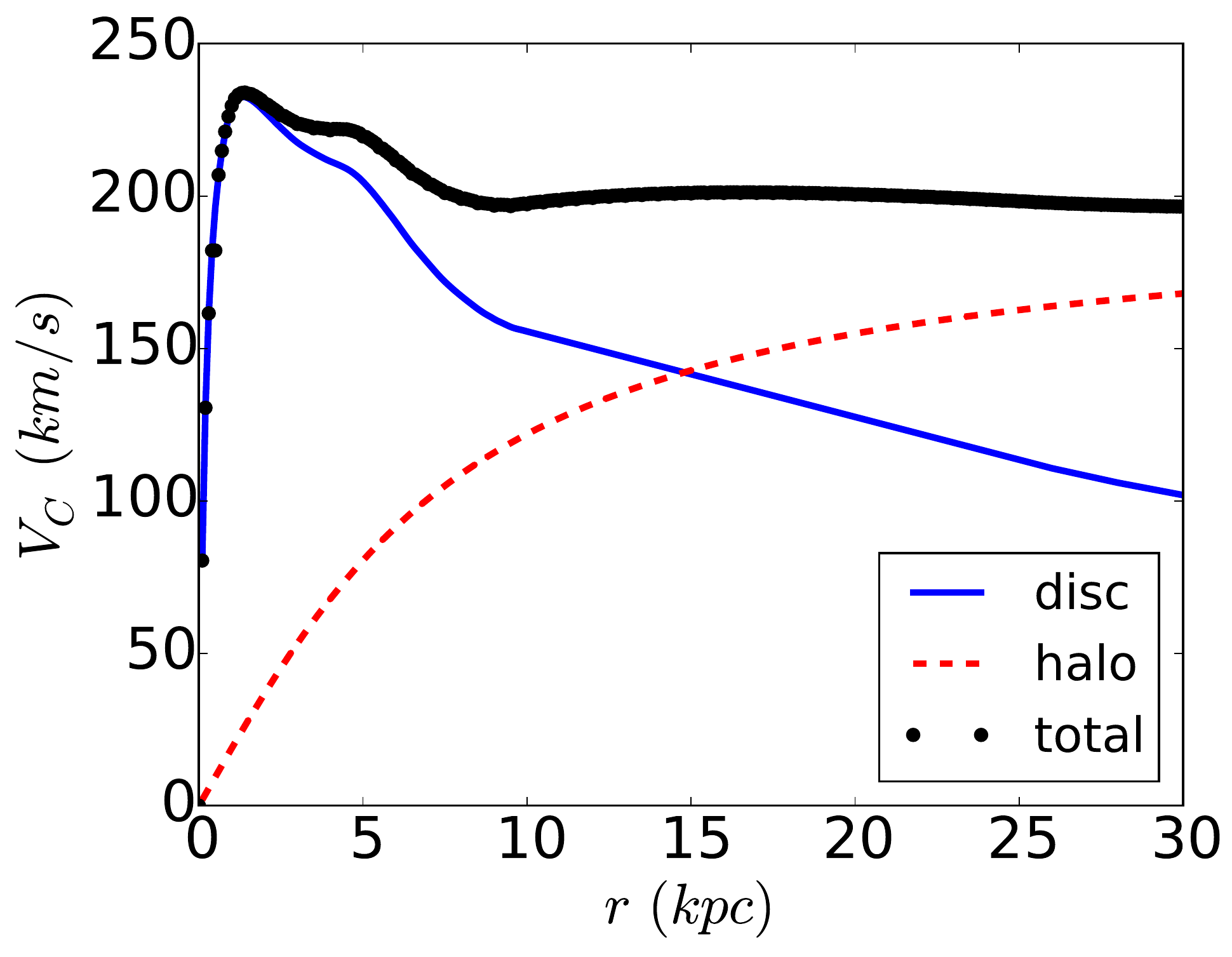}
	\label{fig:rotMD}}
\quad
\subfigure[Sub-Maximal Disc]{%
	\includegraphics[width=0.45\linewidth]{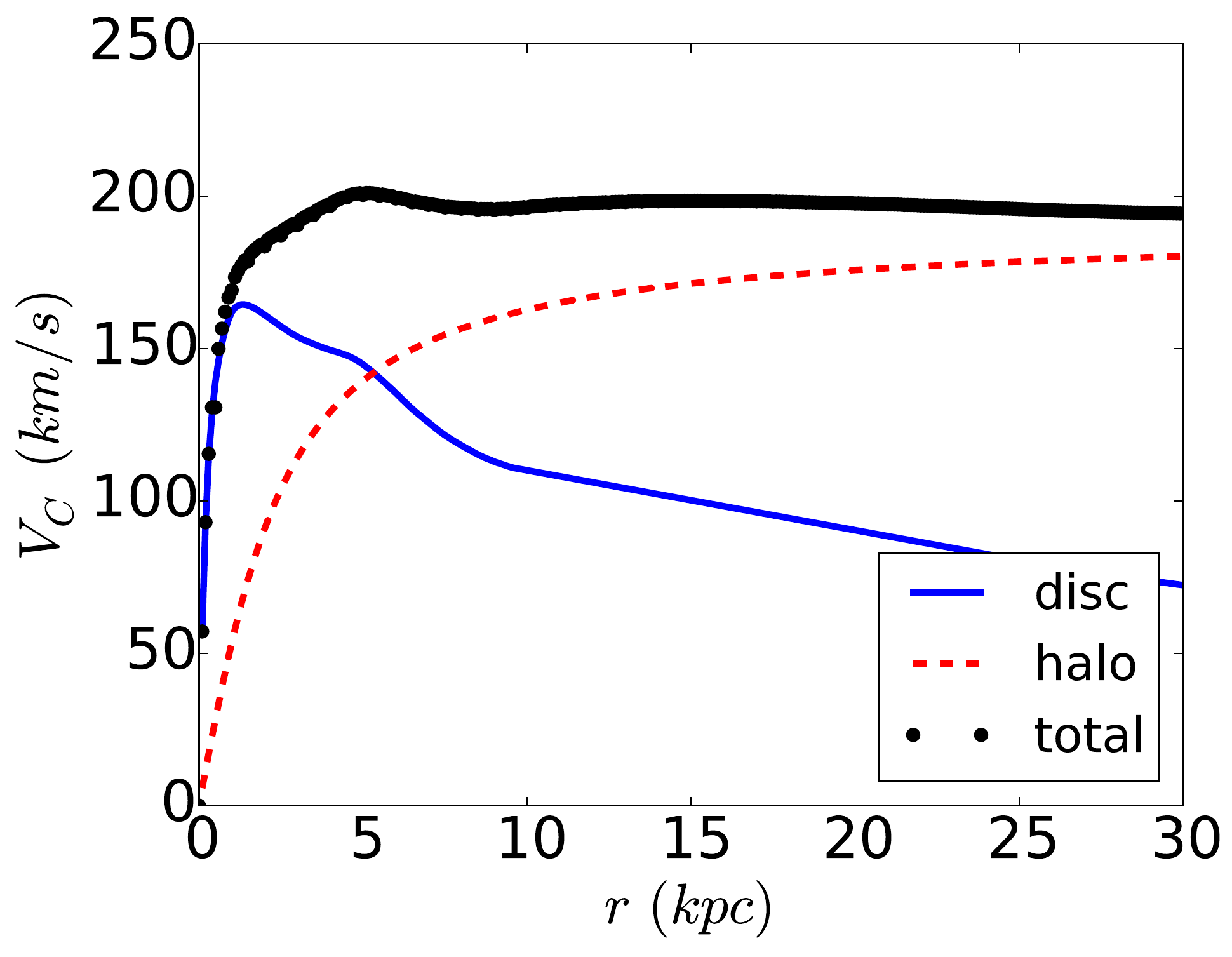}
	\label{fig:rotSMD}}
\quad
\caption{The solid blue line corresponds to the rotation curve of the disc component, and therefore in models without a dark halo represents the total rotation curve. $Left$: Rotation curve for MD model, i.e. models with M/L=0.6 $M_{\odot}$/$L_{\odot}$, where the disc is maximal and a dark matter halo is added such that the $V_{flat}$ of the model matches that predicted by the Tully-Fisher relation. $Right$: Rotation curve for SMD models with a reduced M/L such that the disc is sub-maximal, where a concentrated dark matter halo is added such that the $V_{flat}$ of the model matches that predicted by the Tully-Fisher relation.}
\label{fig:rot}
\end{figure}

\begin{figure}
\centering
\includegraphics[width=0.7\linewidth]{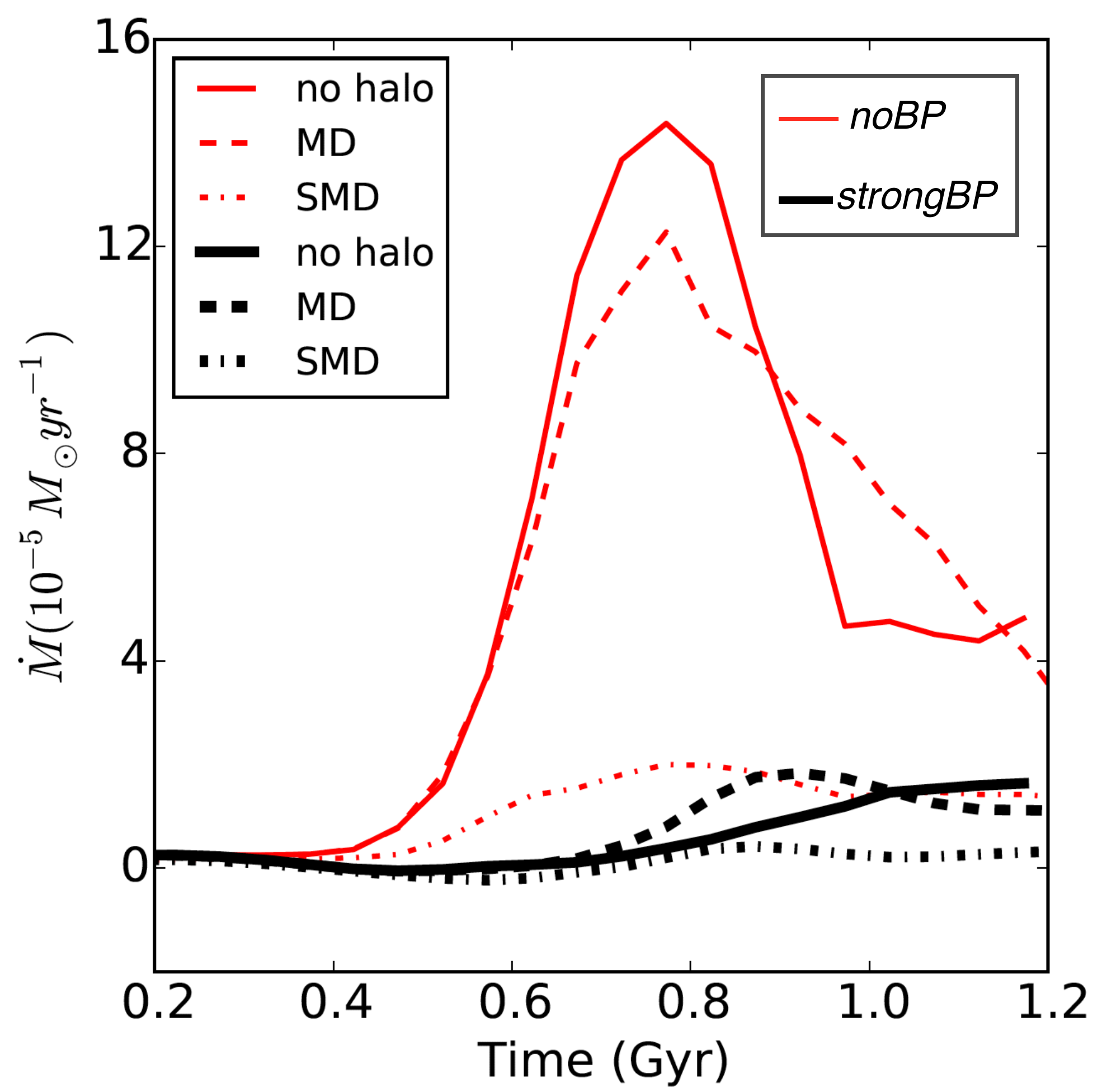}
\caption{Gas inflow rate within 1\,kpc for models with and without a dark matter halo. The thin red lines show the inflow rate for models \emph{noBP} and the thick black lines show the inflow rate for models \emph{strongBP}. Models without a halo, with a maximal disc and a sub maximal disc are shown with solid, dashed and dashed-dotted lines, respectively. We see that there is a general trend for the inflow rate to reduce for more concentrated halos.}
\label{fig:rate_halo_nohalo}
\end{figure}

In Figures \ref{fig:rate} and \ref{fig:tot} we examine how b/p bulges affect the gas inflow rate and total gas mass in the central 1\,kpc. In Figure \ref{fig:rate} we see that in the \emph{weakBP} and \emph{strongBP} models the inflow rate is significantly reduced compared to the \emph{noBP} model. Even after the bar growth period (indicated by the vertical red line), i.e. during the time when the bar rotates with constant strength, we find a considerable difference between cases with and without a b/p. By examining the bottom panel of Figure \ref{fig:rate} -- where we show the relative difference in inflow rate between models \emph{noBP}-\emph{weakBP} (dashed line) and models \emph{noBP} - \emph{strongBP} (dash-dotted line) -- we see that this difference in inflow rate can reach up to two orders of magnitude. This is related to the decreased bar strength in the models with a b/p bulge, as can be seen by comparing Figures \ref{fig:qt} and \ref{fig:rate}.

A direct consequence of the reduced gas inflow rate is of course the reduced amount of gas mass which accumulates in the central kpc for models with a b/p, as can be seen in Figure \ref{fig:tot}. It is worth noting that, as the simulations do not contain self-gravity, the amount of gas which arrives within the central kpc is proportional to the amount of gas in the initial setup. We therefore show the amount of gas within 1\,kpc normalised by the surface density of the gas within 1\,kpc at the start of the the simulation. As can be seen in the bottom panel of Figure \ref{fig:tot}, the gas mass within 1\,kpc is reduced by about a factor of 2 for the \emph{weakBP} model at the time when the difference is maximum, while for the \emph{strongBP} model the gas mass can be reduced by up to a factor 4. 

When examining the inflow rate curve as a function of time in the $noBP$ model (Figure \ref{fig:rate}) we see that the inflow rate first increases, and just before 0.8\,Gyr it starts decreasing (while remaining positive). From test-simulations which we ran over 2.5\,Gyr, we see that the inflow rate continues to decrease with time, despite the fact that the potential is the most non-axisymmetric after 0.9\,Gyr. This presumably occurs due to the fact that, by 0.8\,Gyr, a large fraction of the gas in the bar region has been transported to the centre and thus there is little gas left in the disc to maintain the previous high inflow rate. This hypothesis is strengthened by the fact that the curves of inflow rate for the three models peak at different times, with the models with a small inflow rate (i.e. \emph{weakBP} and \emph{strongBP}) peaking at later times than the curve with the high inflow rate (i.e. \emph{noBP}).  

We further confirm this hypothesis by examining a ``toy model'' of the gas inflow rate for the \emph{noBP} model, shown in Figure \ref{fig:rate_norm}. In this toy model, we assume that the gas inflow rate is proportional to the bar strength (times an arbitrary multiplication constant) $\times$ the gas mass available in the bar region. We see that the shape of the inflow rate in this toy model approximates quite well the inflow rate within 1\,kpc in the full hydrodynamic simulation, thus confirming that the gas inflow rate decreases due to the decreased amount of gas available in the bar region. 

\subsection{The effect of the dark matter halo concentration}
\label{sec:dmflow}
 
By adding a spherical dark matter halo to our models the potential becomes less non-axisymmetric, thus reducing the effect that bars and other non-axisymmetries have on the gas dynamics (see Figure \ref{fig:qt}).
To explore how much this would affect our results we examine two additional sets of models, with different dark matter halo concentrations. The first set of models, the maximal disc models (MD) have low concentration dark matter halos. The second set of models have sub-maximal discs (SMD) and more concentrated dark matter halos (see Figure \ref{fig:rot} and Section \ref{sec:5models}).

We see in Figure \ref{fig:rate_halo_nohalo} that, as expected, the rate of gas inflow tends to be reduced in the cases with the dark matter halos, since the non-axisymmetric forces are weakened. Additionally, we see that the difference between the cases with and without a b/p is also reduced (this can be best seen by examining the models with a concentrated dark matter halo, labeled SMD in Figure \ref{fig:rate_halo_nohalo}). For the maximal disc setups (labeled MD in Figure \ref{fig:rate_halo_nohalo}) the difference between models with and without a b/p, remains significant. Therefore, for the M/L employed in this study (0.6 $M_{\odot}$/$L_{\odot}$), which leads to a maximal disc, the dark matter halo does not have an important effect on the gas inflow, and therefore the effect of the b/p bulge on the gas inflow rate remains significant. 
In general, for massive high surface brightness galaxies -- which are expected to have maximum discs --the effect of the b/p bulge will be significant.


\section{Conclusions}
\label{sec:5summary}

In \cite{Fragkoudietal2015} we found that b/p bulges have a significant effect on models of barred galaxies, i.e. they reduce the bar strength of the model, and alter the orbital structure of the galaxy. Since the properties of periodic orbits in barred galaxies are tightly related to the gas flows \citep{Athanassoula1992a,Athanassoula1992b}, this already hinted at the effect that b/ps would have on the gas inflow. To explore this in detail, here we examine the effects of b/p bulges on the gaseous inflow to the central kiloparsec of barred galaxies, using hydrodynamic simulations. These were run using realistic potentials of a barred galaxy, where the only difference between the models is the presence or not of a b/p bulge. 

We find that both for the weak and strong b/p setups, the rate of gas inflow to the central kiloparsec is reduced by 1-2 orders of magnitude. This leads to a significant reduction (of up to a factor 4) of the total gas mass present inside this radius. 

We also examined the effect of adding a dark matter halo to the models, in order to check whether adding an axisymmetric component would significantly alter the conclusions. For models with maximal discs, the presence of a dark matter halo does not significantly change the results, since the baryonic matter dominates in the central regions. For models which have a sub-maximal disc  -- and therefore a more concentrated dark matter halo -- the gas inflow to the central regions of the galaxy is reduced for all models, i.e. those with and without b/ps.

The reduction in gas inflow rate due to the presence of a b/p bulge, which will take place after the bar has formed, could have a number of effects on the evolution of the galaxy. One potential effect is the reduction of the fuel reservoir for AGN activity; indeed, this could be a step towards explaining results in the literature which do not find a significant correlation between bars and AGN activity (e.g. \citealt{Leeetal2012,Cisternasetal2013,Cheungetal2015}, but see also \citealt{Knapenetal2000,Laineetal2002,Laurikainenetal2004,OhOhYi2012}), although it is of course important to remember that other processes can affect this, such as physical processes occurring near the black hole. 

Additionally, the lower gas density in the central kiloparsec could also have an impact on the mass of discy bulges formed via the bar instability, on the amount of central star formation, stellar feedback and outflows, and on the overall secular evolution of disc galaxies. 

\section*{Acknowledgements}
We thank Francesca Iannuzzi and Dimitri Gadotti for useful comments on the manuscript. We acknowledge financial support to the DAGAL network from the People Programme (Marie Curie Actions) of the European Union's Seventh Framework Programme FP7/2007-2013/ under REA grant agreement number PITN-GA-2011-289313. EA and AB also acknowledge financial support from the CNES (Centre National d'Etudes Spatiales - France). This work was granted access to the HPC resources of Aix-Marseille Université financed by the project Equip@Meso (ANR-10-EQPX-29-01) of the program ``Investissements d'Avenir'' supervised by the Agence Nationale pour la Recherche.

\bibliographystyle{mn2e}
\bibliography{References}

\begin{thebibliography}{}

\bibitem[\protect\citeauthoryear{{Aguerri}, {M{\'e}ndez-Abreu} \&
  {Corsini}}{{Aguerri} et~al.}{2009}]{Aguerrietal2009}
{Aguerri} J.~A.~L.,  {M{\'e}ndez-Abreu} J.,    {Corsini} E.~M.,  2009, \aap,
  495, 491

\bibitem[\protect\citeauthoryear{{Athanassoula}}{{Athanassoula}}{1992a}]{Athanassoula1992a}
{Athanassoula} E.,  1992a, \mnras, 259, 328

\bibitem[\protect\citeauthoryear{{Athanassoula}}{{Athanassoula}}{1992b}]{Athanassoula1992b}
{Athanassoula} E.,  1992b, \mnras, 259, 345

\bibitem[\protect\citeauthoryear{{Athanassoula}}{{Athanassoula}}{2005}]{Athanassoula2005}
{Athanassoula} E.,  2005, \mnras, 358, 1477

\bibitem[\protect\citeauthoryear{{Athanassoula}}{{Athanassoula}}{2008}]{Athanassoula2008}
{Athanassoula} E.,  2008, ArXiv e-prints

\bibitem[\protect\citeauthoryear{{Athanassoula}}{{Athanassoula}}{2016}]{Athanassoula2016}
{Athanassoula} E.,  2016, Galactic Bulges, 418, 391

\bibitem[\protect\citeauthoryear{{Athanassoula} \& {Bureau}}{{Athanassoula} \&
  {Bureau}}{1999}]{Athanassoula&Bureau1999}
{Athanassoula} E.,  {Bureau} M.,  1999, \apj, 522, 699

\bibitem[\protect\citeauthoryear{{Athanassoula}, {Laurikainen}, {Salo} \&
  {Bosma}}{{Athanassoula} et~al.}{2015}]{Athanassoulaetal2015}
{Athanassoula} E.,  {Laurikainen} E.,  {Salo} H.,    {Bosma} A.,  2015, \mnras,
  454, 3843

\bibitem[\protect\citeauthoryear{{Athanassoula}, {Machado} \&
  {Rodionov}}{{Athanassoula} et~al.}{2013}]{Athanassoulaetal2013}
{Athanassoula} E.,  {Machado} R.~E.~G.,    {Rodionov} S.~A.,  2013, \mnras,
  429, 1949

\bibitem[\protect\citeauthoryear{{Binney}}{{Binney}}{1981}]{Binney1981}
{Binney} J.,  1981, \mnras, 196, 455

\bibitem[\protect\citeauthoryear{{Bureau} \& {Athanassoula}}{{Bureau} \&
  {Athanassoula}}{1999}]{Bureau&Athanassoula1999}
{Bureau} M.,  {Athanassoula} E.,  1999, \apj, 522, 686

\bibitem[\protect\citeauthoryear{{Bureau} \& {Athanassoula}}{{Bureau} \&
  {Athanassoula}}{2005}]{Bureau&Athanassoula2005}
{Bureau} M.,  {Athanassoula} E.,  2005, \apj, 626, 159

\bibitem[\protect\citeauthoryear{{Buta} \& {Block}}{{Buta} \&
  {Block}}{2001}]{ButaBlock2001}
{Buta} R.,  {Block} D.~L.,  2001, \apj, 550, 243

\bibitem[\protect\citeauthoryear{{Buta} et~al.,}{{Buta}
  et~al.}{2010}]{Butaetal2010}
{Buta} R.~J.,  et~al., 2010, \apjs, 190, 147

\bibitem[\protect\citeauthoryear{{Buta} et~al.,}{{Buta}
  et~al.}{2015}]{Butaetal2015}
{Buta} R.~J.,  et~al., 2015, \apjs, 217, 32

\bibitem[\protect\citeauthoryear{{Cheung} et~al.,}{{Cheung}
  et~al.}{2015}]{Cheungetal2015}
{Cheung} E.,  et~al., 2015, \mnras, 447, 506

\bibitem[\protect\citeauthoryear{{Chung} \& {Bureau}}{{Chung} \&
  {Bureau}}{2004}]{Chung&Bureau2004}
{Chung} A.,  {Bureau} M.,  2004, \aj, 127, 3192

\bibitem[\protect\citeauthoryear{{Cisternas} et~al.,}{{Cisternas}
  et~al.}{2013}]{Cisternasetal2013}
{Cisternas} M.,  et~al., 2013, \apj, 776, 50

\bibitem[\protect\citeauthoryear{{Coelho} \& {Gadotti}}{{Coelho} \&
  {Gadotti}}{2011}]{CoelhoGadotti2011}
{Coelho} P.,  {Gadotti} D.~A.,  2011, \apjl, 743, L13

\bibitem[\protect\citeauthoryear{{Combes}, {Debbasch}, {Friedli} \&
  {Pfenniger}}{{Combes} et~al.}{1990}]{Combesetal1990}
{Combes} F.,  {Debbasch} F.,  {Friedli} D.,    {Pfenniger} D.,  1990, \aap,
  233, 82

\bibitem[\protect\citeauthoryear{{Combes} \& {Sanders}}{{Combes} \&
  {Sanders}}{1981}]{CombesSanders1981}
{Combes} F.,  {Sanders} R.~H.,  1981, \aap, 96, 164

\bibitem[\protect\citeauthoryear{{Di Matteo}, {Haywood}, {G{\'o}mez}, {van
  Damme}, {Combes}, {Hall{\'e}}, {Semelin}, {Lehnert} \& {Katz}}{{Di Matteo}
  et~al.}{2014}]{DiMatteoetal2014}
{Di Matteo} P.,  {Haywood} M.,  {G{\'o}mez} A.,  {van Damme} L.,  {Combes} F.,
  {Hall{\'e}} A.,  {Semelin} B.,  {Lehnert} M.~D.,    {Katz} D.,  2014, \aap,
  567, A122

\bibitem[\protect\citeauthoryear{{Ellison}, {Nair}, {Patton}, {Scudder},
  {Mendel} \& {Simard}}{{Ellison} et~al.}{2011}]{Ellisonetal2011}
{Ellison} S.~L.,  {Nair} P.,  {Patton} D.~R.,  {Scudder} J.~M.,  {Mendel}
  J.~T.,    {Simard} L.,  2011, \mnras, 416, 2182

\bibitem[\protect\citeauthoryear{{Emsellem}, {Renaud}, {Bournaud}, {Elmegreen},
  {Combes} \& {Gabor}}{{Emsellem} et~al.}{2014}]{Emsellemetal2014}
{Emsellem} E.,  {Renaud} F.,  {Bournaud} F.,  {Elmegreen} B.,  {Combes} F.,
  {Gabor} J.,  2014, ArXiv e-prints

\bibitem[\protect\citeauthoryear{{Eskridge} et~al.,}{{Eskridge}
  et~al.}{2000}]{Eskridgeetal2000}
{Eskridge} P.~B.,  et~al., 2000, \aj, 119, 536

\bibitem[\protect\citeauthoryear{{Fragkoudi}, {Athanassoula}, {Bosma} \&
  {Iannuzzi}}{{Fragkoudi} et~al.}{2015}]{Fragkoudietal2015}
{Fragkoudi} F.,  {Athanassoula} E.,  {Bosma} A.,    {Iannuzzi} F.,  2015,
  \mnras, 450, 229

\bibitem[\protect\citeauthoryear{{Gadotti}}{{Gadotti}}{2009}]{Gadotti2009}
{Gadotti} D.~A.,  2009, \mnras, 393, 1531

\bibitem[\protect\citeauthoryear{{Howard} et~al.,}{{Howard}
  et~al.}{2009}]{Howardetal2009}
{Howard} C.~D.,  et~al., 2009, \apjl, 702, L153

\bibitem[\protect\citeauthoryear{{Kim}, {Seo}, {Stone}, {Yoon} \&
  {Teuben}}{{Kim} et~al.}{2012}]{Kimetal2012}
{Kim} W.-T.,  {Seo} W.-Y.,  {Stone} J.~M.,  {Yoon} D.,    {Teuben} P.~J.,
  2012, \apj, 747, 60

\bibitem[\protect\citeauthoryear{{Knapen}, {Shlosman} \& {Peletier}}{{Knapen}
  et~al.}{2000}]{Knapenetal2000}
{Knapen} J.~H.,  {Shlosman} I.,    {Peletier} R.~F.,  2000, \apj, 529, 93

\bibitem[\protect\citeauthoryear{{Kormendy} \& {Kennicutt} Jr.}{{Kormendy} \&
  {Kennicutt}}{2004}]{KormendyKennicutt2004}
{Kormendy} J.,  {Kennicutt} Jr. R.~C.,  2004, \araa, 42, 603

\bibitem[\protect\citeauthoryear{{Laine}, {Shlosman}, {Knapen} \&
  {Peletier}}{{Laine} et~al.}{2002}]{Laineetal2002}
{Laine} S.,  {Shlosman} I.,  {Knapen} J.~H.,    {Peletier} R.~F.,  2002, \apj,
  567, 97

\bibitem[\protect\citeauthoryear{{Laurikainen}, {Salo}, {Athanassoula}, {Bosma}
  \& {Herrera-Endoqui}}{{Laurikainen} et~al.}{2014}]{Laurikainenetal2014}
{Laurikainen} E.,  {Salo} H.,  {Athanassoula} E.,  {Bosma} A.,
  {Herrera-Endoqui} M.,  2014, \mnras, 444, L80

\bibitem[\protect\citeauthoryear{{Laurikainen}, {Salo} \& {Buta}}{{Laurikainen}
  et~al.}{2004}]{Laurikainenetal2004}
{Laurikainen} E.,  {Salo} H.,    {Buta} R.,  2004, \apj, 607, 103

\bibitem[\protect\citeauthoryear{{Lee}, {Woo}, {Lee}, {Hwang}, {Lee}, {Sohn} \&
  {Lee}}{{Lee} et~al.}{2012}]{Leeetal2012}
{Lee} G.-H.,  {Woo} J.-H.,  {Lee} M.~G.,  {Hwang} H.~S.,  {Lee} J.~C.,  {Sohn}
  J.,    {Lee} J.~H.,  2012, \apj, 750, 141

\bibitem[\protect\citeauthoryear{{Li}, {Shen} \& {Kim}}{{Li}
  et~al.}{2015}]{Lietal2015}
{Li} Z.,  {Shen} J.,    {Kim} W.-T.,  2015, \apj, 806, 150

\bibitem[\protect\citeauthoryear{{L{\"u}tticke}, {Dettmar} \&
  {Pohlen}}{{L{\"u}tticke} et~al.}{2000}]{Luttickeetal2000}
{L{\"u}tticke} R.,  {Dettmar} R.-J.,    {Pohlen} M.,  2000, \aaps, 145, 405

\bibitem[\protect\citeauthoryear{{Martinez-Valpuesta}, {Shlosman} \&
  {Heller}}{{Martinez-Valpuesta} et~al.}{2006}]{MartinezValpuestaetal2006}
{Martinez-Valpuesta} I.,  {Shlosman} I.,    {Heller} C.,  2006, \apj, 637, 214

\bibitem[\protect\citeauthoryear{{McGaugh}, {Schombert}, {Bothun} \& {de
  Blok}}{{McGaugh} et~al.}{2000}]{McGaughetal2000}
{McGaugh} S.~S.,  {Schombert} J.~M.,  {Bothun} G.~D.,    {de Blok} W.~J.~G.,
  2000, \apjl, 533, L99

\bibitem[\protect\citeauthoryear{{Meidt} et~al.,}{{Meidt}
  et~al.}{2014}]{Meidtetal2014}
{Meidt} S.~E.,  et~al., 2014, \apj, 788, 144

\bibitem[\protect\citeauthoryear{{Men{\'e}ndez-Delmestre}, {Sheth},
  {Schinnerer}, {Jarrett} \& {Scoville}}{{Men{\'e}ndez-Delmestre}
  et~al.}{2007}]{Menendezetal2007}
{Men{\'e}ndez-Delmestre} K.,  {Sheth} K.,  {Schinnerer} E.,  {Jarrett} T.~H.,
   {Scoville} N.~Z.,  2007, \apj, 657, 790

\bibitem[\protect\citeauthoryear{{Ness} et~al.,}{{Ness}
  et~al.}{2012}]{Nessetal2012}
{Ness} M.,  et~al., 2012, \apj, 756, 22

\bibitem[\protect\citeauthoryear{{Ness} et~al.,}{{Ness}
  et~al.}{2013}]{Nessetal2013a}
{Ness} M.,  et~al., 2013, \mnras, 430, 836

\bibitem[\protect\citeauthoryear{{Ness} \& {Lang}}{{Ness} \&
  {Lang}}{2016}]{NessLang2016}
{Ness} M.,  {Lang} D.,  2016, ArXiv e-prints

\bibitem[\protect\citeauthoryear{{Oh}, {Oh} \& {Yi}}{{Oh}
  et~al.}{2012}]{OhOhYi2012}
{Oh} S.,  {Oh} K.,    {Yi} S.~K.,  2012, \apjs, 198, 4

\bibitem[\protect\citeauthoryear{{Patsis} \& {Athanassoula}}{{Patsis} \&
  {Athanassoula}}{2000}]{PatsisAthanassoula2000}
{Patsis} P.~A.,  {Athanassoula} E.,  2000, \aap, 358, 45

\bibitem[\protect\citeauthoryear{{Pfenniger} \& {Friedli}}{{Pfenniger} \&
  {Friedli}}{1991}]{PfennigerFriedli1991}
{Pfenniger} D.,  {Friedli} D.,  1991, \aap, 252, 75

\bibitem[\protect\citeauthoryear{{Piner}, {Stone} \& {Teuben}}{{Piner}
  et~al.}{1995}]{Pineretal1995}
{Piner} B.~G.,  {Stone} J.~M.,    {Teuben} P.~J.,  1995, \apj, 449, 508

\bibitem[\protect\citeauthoryear{{Querejeta} et~al.,}{{Querejeta}
  et~al.}{2015}]{Querejetaetal2015}
{Querejeta} M.,  et~al., 2015, \apjs, 219, 5

\bibitem[\protect\citeauthoryear{{Regan} \& {Teuben}}{{Regan} \&
  {Teuben}}{2004}]{ReganTeuben2004}
{Regan} M.~W.,  {Teuben} P.~J.,  2004, \apj, 600, 595

\bibitem[\protect\citeauthoryear{{R{\"o}ck}, {Vazdekis}, {Peletier}, {Knapen}
  \& {Falc{\'o}n-Barroso}}{{R{\"o}ck} et~al.}{2015}]{Roecketal2015}
{R{\"o}ck} B.,  {Vazdekis} A.,  {Peletier} R.~F.,  {Knapen} J.~H.,
  {Falc{\'o}n-Barroso} J.,  2015, \mnras, 449, 2853

\bibitem[\protect\citeauthoryear{{Sackett}}{{Sackett}}{1997}]{Sackett1997}
{Sackett} P.~D.,  1997, \apj, 483, 103

\bibitem[\protect\citeauthoryear{{Shen}, {Rich}, {Kormendy}, {Howard}, {De
  Propris} \& {Kunder}}{{Shen} et~al.}{2010}]{Shenetal2010}
{Shen} J.,  {Rich} R.~M.,  {Kormendy} J.,  {Howard} C.~D.,  {De Propris} R.,
  {Kunder} A.,  2010, \apjl, 720, L72

\bibitem[\protect\citeauthoryear{{Sheth} et~al.,}{{Sheth}
  et~al.}{2010}]{Sheth2010_s4g}
{Sheth} K.,  et~al., 2010, \pasp, 122, 1397

\bibitem[\protect\citeauthoryear{{Shlosman}, {Begelman} \& {Frank}}{{Shlosman}
  et~al.}{1990}]{Shlosmanetal1990}
{Shlosman} I.,  {Begelman} M.~C.,    {Frank} J.,  1990, \nat, 345, 679

\bibitem[\protect\citeauthoryear{{Skokos}, {Patsis} \& {Athanassoula}}{{Skokos}
  et~al.}{2002}]{Skokosetal2002}
{Skokos} C.,  {Patsis} P.~A.,    {Athanassoula} E.,  2002, \mnras, 333, 847

\bibitem[\protect\citeauthoryear{{Teyssier}}{{Teyssier}}{2002}]{Teyssier2002}
{Teyssier} R.,  2002, \aap, 385, 337

\bibitem[\protect\citeauthoryear{{Wegg} \& {Gerhard}}{{Wegg} \&
  {Gerhard}}{2013}]{WeggGerhard2013}
{Wegg} C.,  {Gerhard} O.,  2013, \mnras, 435, 1874

\bibitem[\protect\citeauthoryear{{Weiland} et~al.,}{{Weiland}
  et~al.}{1994}]{Weilandetal1994}
{Weiland} J.~L.,  et~al., 1994, \apjl, 425, L81

\bibitem[\protect\citeauthoryear{{Zaritsky} et~al.,}{{Zaritsky}
  et~al.}{2014}]{Zaritskyetal2014}
{Zaritsky} D.,  et~al., 2014, \aj, 147, 134

\end{thebibliography}



\label{lastpage}

\end{document}